\documentclass{jaums}

\theoremstyle{thmit} 
\newtheorem{thm}{Theorem}[section]

\theoremstyle{thmrm} 

\newtheorem*{oldproof}{Proof}

\newcommand{\R}{\mbox{\rm I \hspace{-0.9em} R}}
\def \part {\partial}
\def \la {\lambda }
\def \ba {\begin{array}}
\def \ea {\end{array}}

\title{Binary Constrained Flows and Separation of Variables for Soliton Equations 
}
\author{Wen-Xiu Ma}
\address{Department of Mathematics
\\City University of Hong Kong\\
Kowloon, Hong Kong, China\\
E-mail:mawx\@cityu.edu.hk}
\author{Yunbo Zeng}
\address{Department of Mathematical Sciences\\
Tsinghua University\\
Beijing 100084, China\\
E-mail:yzeng\@tsinghua.edu.cn}

\date{31 May 2000}

\begin{document}

\maketitle

\centerline{Dedicated to Martin Kruskal's seventy-fifth birthday}

\begin{abstract}
In contrast to mono-constrained flows with $N$ degrees of freedom, 
binary constrained flows of soliton equations, admitting 
$2\times 2$ Lax matrices, have $2N$  degrees of freedom.
By means of the existing method,
Lax matrices only yield
the first $N$ pairs of canonical separated variables. 
An approach for  
constructing the second $N$ pairs of canonical separated variables with  
additional $N$ separated equations is introduced.
The Jacobi inversion problems 
for binary constrained flows are then established. 
Finally, 
the separability of  binary constrained flows
together with the factorization 
 of soliton equations by the spatial and temporal
binary constrained flows leads to the 
Jacobi inversion problems for soliton equations.

\end{abstract}


\section{Introduction}

The separation of variables is one of the most universal methods for 
solving completely integrable models, both classical and quantum. 
If a finite-dimensional integrable Hamiltonian system (FDIHS) with $m$ degrees of freedom
has $m$ functionally independent and involutive 
integrals of motion $P_i,
\ 1\le i\le m$, 
the separation of variables 
\cite{Sklyanin-PTP1995,Kuznetsov-JMP1992} means to construct $m$ pairs of canonical variables    
\begin{equation}
\{u_k, u_l\}=\{v_k, v_l\}=0,\  \{v_k, u_l\}=\delta_{kl},\
 1\le  k,l\le m, \label{1.1}
\end{equation}  
and $m$ separated equations
\begin{equation} f_k(u_k, v_k, P_1,...,P_{m})=0, \  
1\le k\le m.\label{1.2}
\end{equation}
Such pairs of variables are called  canonical separated variables.

For a FDIHS admitting a $2\times 2$ Lax matrix, 
there exists a general method to construct 
canonical separated variables based on the Lax matrix
(for example, see \cite{Sklyanin-PTP1995}-\cite{Zeng-JPA1997}). The corresponding separated equations enable us to express 
the generating function of canonical transformation in a completely 
separated form as an Abelian integral on the associated invariant spectral curve. The resulting linearizing map is essentially the 
Abel map to the Jacobi variety of the spectral curve, thereby providing 
a link with the algebro-geometric linearization method \cite{8.}.
An important feature of the separation of variables for a FDIHS is that 
the number of pairs of canonical separated variables must be equal to the 
number of degrees of freedom. However, in some cases, 
it is found that the existing method may not yield enough pairs of 
canonical separated variables. 
It has been a challenging problem 
\cite{Sklyanin-PTP1995} how to construct additional canonical separated variables
which are required for separation of variables.

Binary constrained flows of soliton hierarchies,  
recently attracting much attention (for example, see \cite{MaS-PLA1994}-\cite{MaF-book1996}), 
are such specific cases, which need to be handled by a different approach.
The degree of freedom of binary constrained flows admitting 
$2\times 2$ Lax matrices is $2N$. 
By using the existing method \cite{Sklyanin-PTP1995,Kuznetsov-JMP1992}, 
the Lax matrices allow to directly construct the first $N$ 
pairs of canonical separated variables $u_1,...,u_N$ and $v_1,...,v_N$. 
In this report, we would like to 
show an approach for determining the second $N$ pairs of canonical separated variables 
and $N$ additional separated equations for binary constrained flows. 
The crucial point is to construct a new set of generating functions $\widetilde B(\lambda )$ and 
$\widetilde A(\lambda )$ 
defining $u_{N+1},...,u_{2N}$ by the set of zeros of $\widetilde B(\lambda )$ 
and 
$v_{N+k}=\widetilde A(u_{N+k}),\ 1\le k\le N.$
To keep the canonical conditions (\ref{1.1}) 
and obtain the separated equations (\ref{1.2}), it is found that 
certain commutator relations need to be imposed on $\widetilde B(\lambda )$ and $
\widetilde A(\lambda )$, and $\widetilde A(\lambda )$ has some link with the 
common generating function of integrals of motion of binary constrained flows, 
which also provides a clue to construct the $\widetilde B(\lambda )$ and $\widetilde 
A(\lambda )$. 
Having analyzed the separation of variables,
the Jacobi inversion problems can be naturally presented for binary constrained 
flows. 

The separation of variables for soliton equations consists of  
two steps of separation of variables \cite{Zeng-JPA1997}. The first step is to factorize 
$1+1$ dimensional soliton equations into two commuting spatial and 
temporal FDIHSs resulted from the spatial and temporal binary constrained flows. 
The second step is to analyze the separation of variables for the 
spatial and temporal binary constrained flows to produce 
their Jacobi inversion problems. 
Finally, combining the factorization of soliton 
equations with the Jacobi inversion problems for the spatial and temporal
binary constrained flows 
enables us to establish the Jacobi inversion problems for soliton 
equations. We will use the AKNS equations \cite{AblowitzKNS-SAM1974} 
to illustrate the whole process. Of course,
the approach adopted can be applied to the whole AKNS hierarchy and 
other soliton hierarchies. 

\section{Separation of variables for binary constrained flows}

\label{approachsection}

Let us first describe binary constrained flows admitting $2\times 2$ Lax matrices,
and then show an approach
for constructing $2N$ pairs of canonical separated variables.

Assume that 
a soliton hierarchy
\begin{equation} 
u_{t_n}=K_n(u)=J\frac{\delta \tilde {H}_n}{\delta u},\ 
u=(u_1,... ,u_q)^T,\ n\ge 0,
\label{sh}
\end{equation} 
where $J$ is a Hamiltonian operator,
is determined by  
a spectral problem 
\begin{equation}
\phi_x=U\phi = U(u,\lambda)\phi,\
U=(U_{ij})_{2\times 2},\  \phi=(\phi_1,\phi_2)^T,
\label{gsp}
\end{equation}
and the associated spectral problems 
\[
\phi_{t_n}=V^{(n)}\phi = V^{(n)}(u,u_x,... ;\lambda)\phi,
\ V^{(n)}=(V^{(n)}_{ij})_{2\times 2}.  
\]
The compatability conditions of the adjoint spectral problem
\begin{equation}
\psi _x=-U^T(u,\la )\psi, \ \psi=(\psi_1,\psi_2)^T
\end{equation}
and the adjoint associated spectral problems 
\[
\psi _{t_n}=-V^{(n)T}\psi =-V^{(n)T}(u,u_x,... ;\la )\psi 
\] 
still give rise to the same soliton hierarchy (\ref{sh}).

Upon introducing $N$ distinct eigenvalues $\la _1,... ,\la _N$,
we have the spatial system
\begin{equation} 
\phi^{(j )}_x=U(u,\la _j )\phi^{(j )}, \ \psi^{(j )}_x=-U^T(u,\la _j )\psi^{(j )},
\label{xpart}
\end{equation} 
where $\phi^{(j)}=(\phi_{1j},\phi_{2j})^T,\ \psi^{(j)}=(\psi_{1j},\psi_{2j})^T,$
 $1\le j\le N$,
and the temporal system
\begin{equation} 
\phi^{(j )}_t=V^{(n)}(u,\la _j )\phi^{(j )}, \ \psi^{(j )}_t=-V^{(n)T}(u,\la _j )\psi^{(j )},
\label{tpart}
\end{equation} 
where $1\le j\le N$.
Let us take the Bargmann symmetry constraint 
\begin{equation}
K_0=J\sum_{j =1}^N E_j  J\frac {\delta \la _j }{\delta u}=
J\sum_{j =1}^N\psi^{(j )T}\frac {\part U(u,\la _j )}{\part u}\phi^{(j )},
\label{gsy}\end{equation} 
where the $E_j$ are normalized constants,
and suppose that (\ref{gsy}) has an inverse function
\begin{equation} u=f(\xi _1,... ,\xi_q),\ \xi_i=\sum_{j =1}^N 
\psi^{(j )T}
\frac {\part U(u,\la _j )}{\part u_i}\phi^{(j )},\ 1\le i\le q.
 \end{equation} 
Replacing $u$ with $f$ in $N$ replicas of   
(\ref{xpart}) and (\ref{tpart}), we obtained the so-called spatial
constrained flow
\begin{equation} 
\phi^{(j )}_x=U(f,\la _j )\phi^{(j )}, \ \psi^{(j )}_x=-U^T(f,\la _j )\psi^{(j )},
\label{xpartofcf}
\end{equation} 
and 
the so-called temporal constrained flow
\begin{equation} 
\phi^{(j )}_t=V^{(n)}(f,f_x,... ;\la _j )\phi^{(j )}, \ \psi^{(j )}_t=-V^{(n)T}(f,f_x,... ;\la _j )\psi^{(j )}, 
\label{tpartofcf}
\end{equation} 
where $1\le j\le N$.
Now if $\phi_{ij}$ and $\psi_{ij}$ solve two constrained flows, then 
$u=f(\xi_1,... ,\xi_q)$ gives rise to a solution to 
the soliton equation $u_{t_n}=K_n(u)$.
The above manipulation is called binary nonlinearization \cite{MaS-PLA1994,Ma-PA1995}.

It is known that constrained flows (CFs) have natural Lax matrices generated from a solution
\[ M(\la )=\left(\ba {cc} A(\la )& B(\la )\vspace{2mm}\\ C(\la )& -A(\la )\ea \right) \]
 to $M_x=[U,M]$ and $M_{t_n}=[V^{(n)},M]$ (for example, see 
\cite{ZengL-JPA1993,AnotonowiczW-IP1993}).
To determine $2N$ pairs of canonical separated variables for binary CFs,
based on Lax matrices $M(\la  )$, we search for two sets of 
generating functions 
${\overline A}(\la ), {\overline B}(\la )$ and $
{\tilde A}(\la ),{\tilde B}(\la )$ such that 
\begin{equation}\left \{\ba {l} 
\{\overline  B(\la), \overline  B(\mu)\}=
\{\widetilde B(\la), \widetilde B(\mu)\}=
\{\overline  A(\la), \overline  A(\mu)\}=
\{\widetilde A(\la), \widetilde A(\mu)\}=0, \qquad \vspace{2mm}\\
\{\overline  B(\la), \widetilde B(\mu)\}=
\{\overline  B(\la), \widetilde A(\mu)\}=
\{\widetilde B(\la), \overline  A(\mu)\}=
\{\overline  A(\la), \widetilde A(\mu)\}=0,\vspace{2mm}\\
\{\overline  A(\la), \overline  B(\mu)\}=\displaystyle{\frac 
{\overline  B(\mu)-\overline  B(\la)}{\la-\mu}},\ \{\widetilde A(\la), \widetilde B(\mu)\}
=\displaystyle{\frac {\widetilde B(\mu)-\widetilde B(\la)}{\la-\mu}},
\ea \right.\label{cr}
\end{equation}
under the standard Poisson bracket 
\begin{equation} \{F,G\}=\sum_{i=1}^2\sum_{j=1}^N\Bigr(
\frac{\part F}{\part \psi_{ij}}\frac {\part G}{\part \phi_{ij}}
- \frac{\part F}{\part \phi_{ij}}\frac {\part G}{\part \psi_{ij}}\Bigl).
\label{Poissonb}\end{equation} 
Such two sets of generating functions 
can be constructed from Lax matrices $M(\la )$ and a common generating function 
of integrals of motion
for binary CFs.
We expect each pair of generating functions 
can yield $N$ pairs of canonical separated variables, through defining
$u_{1},...,u_{N}$ by the set of zeros of $\overline B(\lambda )$,
$u_{N+1},...,u_{2N}$ by the set of zeros of $\widetilde B(\lambda )$, 
and  
\begin{equation} v_{k}=\overline A(u_{k}),\ v_{N+k}=\widetilde A(u_{N+k}),\ 1\le k\le N,
\label{defofv_kandv_{N+k}}\end{equation}
which will also give us all $2N$ separated equations.
Therefore,
the separation of variables for binary CFs   
becomes the problem to find two sets of generating functions
satisfying the above commutator relations (\ref{cr}).
The whole process will be illustrated by the AKNS equations.
 
\section{Binary constrained flows of the AKNS equations}

Let us start from the AKNS spectral problem 
\begin{equation} 
 \phi_x=U\phi=U(u,\lambda )\phi,\ 
 U=\left( \begin{array}{cc} -\lambda &q\vspace{2mm} \\r&\lambda \end{array}\right),\  
\phi =\left(\begin{array}{c}
\phi_{1}\vspace{2mm}\\
\phi_{2} \end{array}\right)
 ,\  u=\left( \begin{array}{c} q\vspace{2mm}\\ r\end{array}\right),\label{2.1}
\end{equation} 
and take the associated spectral problems
\begin{equation} 
 \phi_{t_n}=V^{(n)}\phi = V^{(n)}(u,\lambda )\phi ,\  V^{(n)}=\sum_{i=0}^{n}
\left( \begin{array}{cc} a_i&b_i\vspace{2mm}\\c_i&-a_i\end{array}\right)\lambda ^{n-i},\label{2.2}\end{equation} 
with $a_i,b_i,c_i$ being defined by
\[\ba {l}
a_{0}=-1,\  b_{0}=c_{0}=0,\  a_1=0,\  b_{1}=q,\ 
c_{1}=r,\  a_2=\frac{1}{2}qr,\ ...,\vspace{2mm}\\
\left(\begin{array}{c} {c_{k+1}}\vspace{2mm}\\{ b_{k+1
}}\end{array}\right)=L \left( \begin{array}{c}c_{k}\vspace{2mm}\\
 b_{k} \end{array}\right), \  
a_{k+1}= \partial ^{-1}(qc_{k+1}-rb_{k+1}),\ k\ge 1,\ea \]
where $L$ is given by
\[ 
L=\frac 12\left( \begin{array}{cc} \partial -2r\partial ^{-1}q&2r\partial ^{-1}r
\vspace{2mm} \\-2q\partial ^{-1}q&-\partial +2q\partial ^{-1}r\end{array}\right).\]
The compatibility conditions of (\ref{2.1}) and (\ref{2.2}) give  
the AKNS hierarchy 
\begin{equation} 
u_{t_n}
=J \left( \begin{array}{c}{c_{n+1}}\vspace{2mm}\\
{b_{n+1}}\end{array}\right)
=J\frac {\delta \tilde {H}_{n+1}}{\delta u},\  J=\left( \begin{array}{cc} 0&-2\\2&0\end{array}\right),\ 
\tilde {H}_{n}=\int \frac{2a_{n+1}}{n+1}dx,
\ n\ge 1, \label{2.3}\end{equation} 
which contains the AKNS equations 
\begin{equation}
q_{t_2}=-\frac 12q_{xx}+q^2r, \  r_{t_2}=\frac 12r_{xx}-r^2q. 
\label{AKNS}\end{equation} 

Introducing $N$ distinct eigenvalues $\la _j,\ 1\le j\le N$, we have 
\begin{equation} \left\{ \begin{array}{l} 
 \Phi_{1x}=-\Lambda  \Phi_{1}+q  \Phi_{2},\
  \Phi_{2x}=r  \Phi_{1}+\Lambda  \Phi_{2}
,\vspace{2mm}\\
  \Psi_{1x}=\Lambda  \Psi_{1}-r  \Psi_{2},\  \Psi_{2x}=-q  \Psi_{1}-\Lambda
  \Psi_{2},\end{array} \right. 
\label{2.5}\end{equation} 
and the Bargmann symmetry constraint reads as 
\begin{equation} 
\frac {\delta H_{1}}{\delta u}-
\sum_{j=1}^{N}\frac {\delta \lambda _{j}}{\delta u}
=\left(\begin{array}{c} {r}\vspace{2mm}\\{q}\end{array}\right)
-\left( \begin{array} {l} 
{\langle  \Psi_1,  \Phi_2\rangle }\vspace{2mm}\\
{ \langle   \Psi_2, \Phi_1\rangle }\end{array}\right)=0,\label{sc}
\end{equation} 
where $\langle \cdot,\cdot\rangle $ denotes the standard inner product of $\R ^{N}$ and 
\[  \Phi_i=(\phi_{i1},...,\phi_{iN})^T, \  \Psi_i=( \psi_{i1},..., \psi_{iN})^T, \ i=1,2,
\  \Lambda =\textrm{diag}(\lambda _1,...,\lambda _N).
 \]
Therefore, the spatial constrained flow (\ref{xpartofcf}) 
is the following $x$-FDIHS \cite{MaS-PLA1994} 
\begin{equation}
  \Phi_{1x}=\frac {\partial  F_1}{\partial   \Psi_1},\   \Phi_{2x}=\frac {\partial  
F_1}{\partial  \Psi_2},\ 
 \Psi_{1x}=-\frac {\partial  F_1}{ \partial   \Phi_1},\ 
  \Psi_{2x}=-\frac {\partial  F_1}{\partial   \Phi_2},\label{2.6}\end{equation} 
with the Hamiltonian
$$F_1=\langle \Lambda  \Psi_2,  \Phi_2\rangle -\langle \Lambda   \Psi_1,   \Phi_1\rangle 
+\langle  \Psi_2,   \Phi_1\rangle \langle  \Psi_1,   \Phi_2\rangle .$$
Under the symmetry constraint (\ref{sc}) and the $x$-FDIHS (\ref{2.6}), 
the binary $t_2$-constrained flow (\ref{tpartofcf})
can be transformed into the following $t_2$-FDIHS
\begin{equation} 
 \Phi_{1t_2}=\frac {\partial  F_2}{ \partial  \Psi_1},\   \Phi_{2t_2}=\frac {\partial  F_2}{\partial  \Psi_2},\ 
 \Psi_{1t_2}=-\frac {\partial  F_2}{\partial  \Phi_1},\ 
  \Psi_{2t_2}=-\frac {\partial  F_2}{\partial   \Phi_2},\label{2.7}\end{equation} 
with the Hamiltonian
\[ \begin{array} {l}
F_2=\langle \Lambda  ^2\Psi_2,  \Phi_2\rangle -\langle \Lambda  ^2  \Psi_1,  \Phi_1\rangle 
+\langle  \Psi_2, \Phi_1\rangle \langle \Lambda   \Psi_1, \Phi_2\rangle +\vspace{2mm}\\
 \langle \Lambda  \Psi_2,  \Phi_1\rangle \langle  \Psi_1, \Phi_2\rangle 
-\frac 12(\langle  \Psi_2,  \Phi_2\rangle -\langle  \Psi_1, \Phi_1\rangle )\langle \Psi_2, \Phi_1\rangle 
\langle  \Psi_1, \Phi_2\rangle . \end{array} 
\]
The Lax matrix $M=\left( \begin{array}{cc} A(\lambda )&B(\lambda )\\C(\lambda )&
-A(\lambda )\end{array}\right)$ for the FDIHSs (\ref{2.6}) and (\ref{2.7}) is given by 
\cite{LiM-CSF2000}
\begin{equation} 
A(\lambda )=-1+\sum_{j=1}^{N}\frac{\psi_{1j}\phi_{1j}- \psi_{2j}\phi_{2j}}
{2(\lambda -\lambda _{j})},\ 
B(\lambda )=\sum_{j=1}^{N}\frac{ \psi_{2j} \phi_{1j}}{\lambda -\lambda _{j}},\ 
C(\lambda )=\sum_{j=1}^{N}\frac{\psi_{1j} \phi_{2j}}{\lambda -\lambda _{j}}. \label{2.8}\end{equation} 
A straightforward calculation yields
\begin{equation} P(\lambda ) 
:= A^2(\lambda )+B(\lambda )C(\lambda )=1+
\sum_{j=1}^{N}[\frac{P_{j}}{\lambda -\lambda _{j}}+\frac{P^2_{N+j}}{(\lambda -\lambda _{j})^2}], \label{2.9}
\end{equation} 
where the $P_j$ and $P_{N+j}$ are $2N$ involutive integrals of motion for (\ref{2.6}) and (\ref{2.7})
\begin{eqnarray} && 
P_j=\frac 12\sum_{k\neq j}\frac{1}{\lambda _j-\lambda _{k}}[(\psi_{1j}\phi_{1j}-\psi_{2j}   \phi_{2j})(\psi_{1k} \phi_{1k}-  \psi_{2k}   \phi_{2k})
\nonumber \\
&& \qquad  + \psi_{2j}\phi_{2j}-\psi_{1j}\phi_{1j}
+4 \psi_{1j}  \phi_{2j} \psi_{2k}   \phi_{1k}],\ 1\le j\le N,\qquad  \\
&& P_{N+j}=\frac 12( \psi_{1j}  \phi_{1j}+ \psi_{2j} \phi_{2j}),\  1\le j\le N.\label{2.10}\end{eqnarray} 
It is easy to verify that
\begin{eqnarray} 
F_1&=&\sum_{j=1}^{N}(\lambda _jP_{j}+P^2_{N+j})-(\sum_{j=1}^{N}\frac {P_{j}}2)^2, \label{F_1andF_2usingP_ia}\\
F_2&=&\sum_{j=1}^{N}(\lambda _j^2P_{j}+2\lambda _jP^2_{N+j})-(\sum_{j=1}^{N}\frac{P_{j}}2)
\sum_{j=1}^{N}(\lambda _jP_{j}+P^2_{N+j})+(\sum_{j=1}^{N}\frac {P_{j}}2)^3. \qquad \label{F_1andF_2usingP_ib}\end{eqnarray} 
With respect to the standard Poisson bracket (\ref{Poissonb}), 
it is found \cite{LiM-CSF2000} that
\begin{equation}\left\{ \ba {l}
\{A(\lambda ), A(\mu)\}=\{B(\lambda ), B(\mu)\}=\{C(\lambda ), C(\mu)\}=0,\vspace{2mm}\\
\{A(\lambda ), B(\mu)\}=\frac 1{\lambda -\mu}[B(\mu)-B(\lambda )],\vspace{2mm}\\ 
\{A(\lambda ), C(\mu)\}=\frac 1{\lambda -\mu}[C(\lambda )-C(\mu)],\ 
\vspace{2mm}\\
\{B(\lambda ), C(\mu)\}=\frac 2{\lambda -\mu}[A(\mu)-A(\lambda )]. \ea \right.
\label{2.12}\end{equation}
Then $\{A^2(\lambda )+B(\lambda )C(\lambda ),  A^2(\mu)+B(\mu)C(\mu)\}=0$ implies that the 
integrals of motion 
$P_j$ and $P_{N+j}$, $1\le j\le N$,
are in involution in pairs. The AKNS equations (\ref{AKNS}) are factorized by the 
$x$-FDIHS (\ref{2.6}) and the $t_2$-FDIHS (\ref{2.7}). Namely,
if $\Phi_1,  \Phi_2, \Psi_1$ and $\Psi_2$ 
solve the $x$-FDIHS (\ref{2.6}) and the $t_2$-FDIHS (\ref{2.7}) simultaneously, 
then $(q, r)$ given by (\ref{sc}) solves the AKNS equations (\ref{AKNS}).

\section{Separation of variables for the AKNS equations}

The commutator relations (\ref{2.12}) and 
a common generating function of integrals of motion 
\[ \frac 12 \sum_{j=1}^N\frac {\phi_{1j}\psi_{1j}+\phi_{2j}\psi_{2j}}{\la -\la _j}=
\sum_{j=1}^N\frac {P_{N+j}}{\la -\la _j}
 \]
enable us to construct 
two sets of generating functions which lead to 
$2N$ pairs of canonical separated variables. 
The required two sets of generating functions for the AKNS equations (\ref{AKNS}) are the following
\begin{eqnarray}
 &&
\overline A(\la )= B(\la )- A(\la )-\widetilde A(\la )=
1+\sum_{j=1}^{N} \frac { (\psi_{2j}-\psi_{1j}) \phi_{1j}}
{\la -\la _j }, \label{overlineAandB} \\&& 
\overline B(\la )= B(\la )-2A(\la )-C(\la )=
2+\sum_{j=1}^{N}\frac{(\psi_{2j}-\psi_{1j})(\phi_{1j}+\phi_{2j})}
{\la -\la _{j}},
 \\&&
\widetilde A(\la )= 
\frac{1}{2}\sum_{j=1}^{N}\frac{\psi_{1j}\phi_{1j}+\psi_{2j}\phi_{2j}}
{\la -\la _{j}},\ \widetilde B(\la )= 
1+\frac{1}{2}\sum_{j=1}^{N}\frac{(\phi_{1j}+\phi_{2j})^2}
{\la -\la _{j}}.\label{widetildeAandB} 
\end{eqnarray}
Let us now introduce $u_k,\,u_{N+k},\ 1\le k\le N,$ by 
\begin{equation}     
\overline B(\la )= 2\frac {\overline R(\la)}{K(\la)},\ 
\widetilde B(\la )=\frac {\widetilde R(\la)}{K(\la)},\label{defofu_kandu_{N+k}}
\end{equation}  
where $\overline R(\la )$, $\widetilde R(\la )$ and $ K(\la )$ read as
\begin{equation} \overline R(\la )= \prod _{k=1}^N(\la -u_{k}),\ 
\widetilde R(\la )= \prod _{k=1}^N(\la -u_{N+k}),\ 
K(\la )= \prod _{k=1}^N(\la -\la _k).
 \ \label{defofu_kandu_{N+k}a}\end{equation}
A direct computation can show the following result. 
\begin{thm} 
Assume that $\lambda _j, 
\phi_{ij}, \psi_{ij}
, \ i=1,2,\  1\le j\le N$, are all real, and
 $u_1,...,u_N$ are single zeros of $\overline B(\lambda )$. Then the variables 
$u_1,...,u_{2N}$ defined by (\ref{defofu_kandu_{N+k}}) and (\ref{defofu_kandu_{N+k}a}),
and the variables 
$v_1,...,v_{2N}$ defined by the corresponding formula 
(\ref{defofv_kandv_{N+k}})
are canonically conjugated, i.e., they satisfy the commutator relations
(\ref{1.1}) with $m=2N$.
\end{thm}

It follows from (\ref{defofu_kandu_{N+k}}) and (\ref{defofu_kandu_{N+k}a}) that
$$ (\psi_{2j}-\psi_{1j})(\phi_{1j}+\phi_{2j})=2\frac{\overline R(\la_j)}{K'(\la_{j})},\ 
 (\phi_{1j}+\phi_{2j})^2=2\frac{\widetilde R(\la_j)}{K'(\la_{j})},\ 1\le j\le N,$$
which leads to  
\begin{equation}
 (\phi_{1j}+\phi_{2j})=\sqrt{\frac{2\widetilde R(\la_j)}{K'(\la_{j})}},\ 
 (\psi_{2j}-\psi_{1j})=\frac {\sqrt{2}\,\overline R(\la_j)}{\sqrt{\widetilde R(\la_j)K'(\la_{j})}},\ 1\le j\le N.\end{equation} 
By substituting $u_k$ into $\overline A(\la )$ 
of (\ref{overlineAandB}), $u_{N+k}$ into $\widetilde A(\la )$
of (\ref{widetildeAandB}) and noting 
\[ (A(\la )-B(\la ))^2-\overline B(\la )B(\la ) =A^2(\la )+B(\la )C(\la )=
P(\la ),
\]
one gets the separated equations
\begin{eqnarray} 
&&v_k=B(u_k)-A(u_k) -\widetilde A (u_k)=\sqrt {P(u_k)}-\widetilde A(u_k)\nonumber \\
&& =
\sqrt {1+\sum_{j=1}^{N}[\frac{P_{j}}{u_k-\la _{j}}+\frac{P^2_{N+j}}{(u_k-\la _{j})^2}]}
  -\sum_{j=1}^{N}\frac{P_{N+j}}{u_k-\la _{j}},\ 1\le  k\le N, \qquad \\ 
&&v_{N+k}= \widetilde A(u_{N+k})
=\sum_{j=1}^{N}\frac{P_{N+j}}{u_{N+k}-\la_{j}}, 
\ 1\le  k\le N. \end{eqnarray}
Replacing $v_k$ by the partial derivative $\frac {\partial S}{\partial u_k}$ of the generating function $S$ of canonical transformation and interpreting the $P_j$ and $P_{N+j}$ 
as integration constants, 
the above separated equations give rise to the Hamilton-Jacobi equations which are 
completely separated and can be integrated to give the completely separated solution for $S$
\begin{eqnarray}&&
S(u_1,...,u_{2N})=\sum_{k=1}^{N}[\int^{u_k}(\sqrt {P(\la )}-\widetilde A(\la ))d\la
+\int^{u_{N+k}}\widetilde A(\la )d\la]\nonumber \\
&&=\sum_{k=1}^{N}[\int^{u_k}\sqrt {P(\la)}d\la-\sum_{j=1}^{N}P_{N+j}\textrm{ln} 
\mid\frac{u_k-\la_j}{u_{N+k}-\la_j}\mid]. \label{defofS}\end{eqnarray}
The linearizing coordinates are then
\begin{equation}\left\{\ba {l} \displaystyle{
Q_j=\frac {\partial S}{\partial P_j}=\frac 12\sum_{k=1}^{N}\int^{u_k}\frac {1}{(\la-\la_j)\sqrt { P(\la)}}d\la , 1\le j\le N, }\vspace{2mm}\\
\displaystyle{Q_{N+j}=\frac {\partial S}{\partial P_{N+j}}
=\sum_{k=1}^{N}[\int^{u_k}\frac {P_{N+j}}{(\la-\la_j)^2\sqrt { P(\la)}}d\la-\textrm{ln} \mid\frac{u_k-\la_j}{u_{N+k}-\la_j}\mid], } \ea \right. \label{Q_iandQ_{N+i}}\qquad \end{equation}
where $\ 1\le j\le N.$
These coordinates $Q_j$ and $Q_{N+j}$, $1\le j\le N$,
constitute the action-angle variables together with 
the $P_j$ and $P_{N+j}$, $1\le j\le N$. 
By using (\ref{F_1andF_2usingP_ia}) and (\ref{F_1andF_2usingP_ib}),
the linear flows induced by the $x$-FDIHS (\ref{2.6}) and the $t_2$-FDIHS  (\ref{2.7})
lead to the Jacobi inversion problem for the $x$-FDIHS (\ref{2.6}) 
\begin{equation}\left\{\ba {l}\displaystyle{
2Q_j=\gamma_{j}+(2\lambda _j-\sum_{k=1}^{N}P_k)x, } \vspace{2mm}\\
\displaystyle{ Q_{N+j}=
\gamma_{N+j}+2P_{N+j}x,}\ea \right. \label{3.8}\end{equation} 
and the Jacobi inversion problem for the $t_2$-FDIHS (\ref{2.7})
\begin{equation} \left\{\begin{array} {l} 
\displaystyle{2Q_j=\bar\gamma_j+[2\lambda _j^2
-\sum_{k=1}^{N}(\lambda _kP_k+\lambda _jP_k+P^2_{N+k})
+\frac 34(\sum_{k=1}^{N}P_k)^2]t_2,}
\vspace{2mm} \\ 
\displaystyle{Q_{N+j} =\bar\gamma_{N+j}+P_{N+j}(4\lambda _j-\sum_{k=1}^{N}P_k)t_2,}
\end{array} \right.
\label{3.9}
\end{equation} 
where $1\le j\le N$, the $Q_j$ and $Q_{N+j}$ are defined by 
(\ref{Q_iandQ_{N+i}}), and $\gamma _j$ and $ \bar \gamma_j,\ 1\le j\le 2N$, are arbitrary constants.

Since the AKNS equations (\ref{AKNS}) are factorized by the $x$-FDIHS (\ref{2.6}) and 
the $t_2$-FDIHS (\ref{2.7}), combining  the Jacobi inversion problems 
(\ref{3.8}) and (\ref{3.9}) together gives rise to the following theorem.
\begin{thm} The AKNS equations (\ref{AKNS})
have the Jacobi inversion problem determined by
\[ \begin{array} {l} 
 \displaystyle{
\sum_{k=1}^{N}\int^{u_k}\frac {1}{(\lambda -\lambda _j)\sqrt {P(\lambda )}}d\lambda \,}
\displaystyle{
=\tilde \gamma_{j}+(2\lambda _j-\sum_{k=1}^{N}P_k)x}\vspace{2mm}\\
\quad \displaystyle{
+[2\lambda _j^2
-\sum_{k=1}^{N}(\lambda _kP_k+\lambda _jP_k+P^2_{N+k})
+\frac 34(\sum_{k=1}^{N}P_k)^2]t_2,}
\vspace{2mm}\\
 \displaystyle{
\sum_{k=1}^{N}[\int^{u_k}\frac {P_{N+j}}{(\lambda -\lambda _j)^2\sqrt { P(\lambda )}}d\lambda -\textrm{ln}\mid\frac {u_k-\lambda _j}{u_{N+k}-\lambda _j}\mid]}
\vspace{2mm}\\
\displaystyle{
=\tilde \gamma_{N+j}+2P_{N+j}x+P_{N+j}(4\lambda _j-\sum_{k=1}^{N}P_k)t_2,}
\end{array} \]
where $1\le j\le N$, and 
$\tilde \gamma _j$ and $\tilde  \gamma_{N+j},\ 1\le j\le N$, are arbitrary constants.
\end{thm}

We remark that 
the above Jacobi inversion problem for
the AKNS equations (\ref{AKNS}) is different from that in \cite{ZengM-JMP1999},
which was generated from another class of 
canonical separated variables for the binary constrained flows (\ref{2.6}) and (\ref{2.7}). 
The above manipulation may also be similarly made
for the whole AKNS hierarchy, and the approach depicted in 
Section \ref{approachsection} can be applied to other soliton hierarchies such as 
the KdV hierarchy and the Kaup-Newell hierarchy.

\newpage 
\noindent {\bf Acknowledgments:}
This work was supported by  the City University of Hong Kong (SRGs: 7000945 and 7001041) and the Research Grants Council of Hong Kong (CERGs: 9040395 and 9040466) and the Chinese Basic Research Project ``Nonlinear Science''. One of the authors (Ma) is also grateful to the organizer 
Nalini Joshi for inviting him to give a talk at 
the workshop Kruskal 2000 in Adelaide.

\end{document}